# Josephson effect in symmetry-mismatched van der Waals hybrid junctions of $NbSe_2$-$Bi_2Sr_2CaCu_2O_{8+x}$ and $NbSe_2$-$Bi_2Sr_{2-x}La_xCuO_{6+y}$

Yuying Zhu[1,2#*], Ziye Liu[#3], Hongtao Yan[4,5], Lin Zhao[4,5], Xingjiang Zhou[4,5], Qi-Kun Xue[1,3,6*], and Ding Zhang[1,2,3*]

[1]Beijing Academy of Quantum Information Sciences, Beijing 100193, China

[2]Hefei National Laboratory, Hefei 230088, China

[3]State Key Laboratory of Low Dimensional Quantum Physics and Department of Physics, Tsinghua University, Beijing 100084, China

[4]Beijing National Laboratory for Condensed Matter Physics, Institute of Physics, Chinese Academy of Sciences, Beijing 100190, China.

[5]School of Physical Sciences, University of Chinese Academy of Sciences, Beijing 100049, China

[6]Southern University of Science and Technology, Shenzhen 518055, China

# These authors contributed equally.

*Email: zhuyy@baqis.ac.cn

qkxue@mail.tsinghua.edu.cn

dingzhang@mail.tsinghua.edu.cn

# Abstract

Coupling superconductors of different symmetries may lead to emergent quantum phenomena but experimental studies on heterojunctions made of lattice-mismatched crystals are scarce. Here, we fabricate Josephson junction by van der Waals stacking $NbSe_2$ flakes on freshly cleaved surfaces of $Bi_2Sr_2CaCu_2O_{8+x}$ or $Bi_2Sr_{2-x}La_xCuO_{6+y}$. We obtain pronounced Josephson currents across the heterojunctions at temperatures below 7 K. The Josephson coupling strength, characterized by the product of the critical Josephson current $I_c$ and the normal state resistance $R_n$, can reach about 0.3 mV, much exceeding the values previously reported in Pb-$Bi_2Sr_2CaCu_2O_{8+x}$ or Nb-$Bi_2Sr_2CaCu_2O_{8+x}$ junctions. Microwave irradiation induces integer Shapiro steps. We also observe the Fraunhofer diffraction pattern by applying an in-plane magnetic field. We compare our results with the theoretical calculations that take into account the coherent tunneling and the order parameter mixing. Our work demonstrates the hybrid junctions as a distinctive platform for addressing pairing symmetry and emergent quantum phenomena.

Josephson tunneling is sensitive to the pairing symmetries of the two participating superconductors. For certain junction orientations, symmetry can suppress the first harmonic term in the current phase relation (CPR) and promote the effect of higher order terms [1-4]. This symmetry consideration predicts that the out-of-plane Josephson coupling between two *d*-wave (specifically, $d_{x^2-y^2}$-wave) superconductors plummets as their relative twist angle is tuned to 45 degrees. At this twist angle, the tunneling problem becomes effectively equivalent to that between a $d_{x^2-y^2}$-wave superconductor and a $d_{xy}$-wave superconductor, i.e., two superconductors with distinctly different symmetries. The fundamental harmonic of the Josephson coupling is expected to vanish. The second harmonic, on the other hand, may become appreciable and can stabilize a superconducting ground state with spontaneous time reversal symmetry breaking (TRSB) [3][5]. Experimental efforts to investigate the theoretical proposals have been carried out by fabricating *c*-axis twisted bicrystals of $Bi_2Sr_2CaCu_2O_{8+x}$ (Bi2212) and $Bi_2Sr_{2-x}La_xCuO_{6+y}$ (Bi2201) [6-9]. Zhao *et al.* reported features such as half-integer Shapiro steps that are consistent with co-tunneling of Cooper pairs [7]. Still, other experiments demonstrated that trapped vortices may be responsible for the same exotic phenomena [10]. Despite this on-going debate, these studies have jointly led to fresh insights into high-temperature superconductivity [11,12] and resulted in functional devices such as high-temperature Josephson diodes [13-15] and superconducting quantum interference devices (SQUIDs) [16].

Theoretically, coupling an *s*-wave superconductor with a *d*-wave one [4] is an alternative route toward the realization of Josephson junctions with dominating higher harmonics in CPR [17,18]. This route is theoretically more straightforward than coupling two *d*-wave superconductors because it is less stringent on the crystalline orientation. Earlier experiments were achieved in junctions between a conventional superconductor of Pb and the high-temperature superconductor of $YBa_2Cu_3O_{7-x}$ (YBCO) [18,19]. This endeavor was soon extended to junctions between $Bi_2Sr_2CaCu_2O_{8+x}$ (Bi2212) and Pb/Nb [20,21]. However, they showed standard temperature dependent behaviors of the Josephson current, standard Fraunhofer diffraction patterns, and integer Shapiro steps in the ac Josephson effect. These results suggest that the Josephson tunneling still originates predominantly from the fundamental harmonic of the CPR. Recently, van der Waal stacking of layered materials has opened a unique pathway toward searching for non-standard Josephson tunneling between different superconductors. This technique allows for atomically flat interfaces between single crystalline materials, preventing the diffusion problem suffered in earlier studies [19]. The crystal symmetries of both participants in the junction can profoundly influence the Josephson coupling [4], which can be further tuned by the twist angle. So far, heterostructures such as Bi2212-Bi2201 [9] and $FeTe_{0.55}Se_{0.45}$-$NbSe_2$ junctions [22] have been reported. However, van der Waals junctions between a conventional superconductor and a cuprate superconductor have not been studied yet.

Here, we address the Josephson effect in *c*-axis van der Waals junctions made of two lattice-mismatched materials: $NbSe_2$ and Bi-based cuprates (Bi2212 and Bi2201). Experimentally, we observe the Josephson effect in both $NbSe_2$-Bi2212 and $NbSe_2$-Bi2201 heterojunctions. The nature of Josephson coupling manifests as integer Shapiro steps in the $I$ - $V$ characteristics when the junctions are irradiated by microwave. Also, applying an in-plane magnetic field induces the Fraunhofer diffraction pattern. The experimental results are compared to our theoretical calculations that take into account the tunneling coherence and the mixing of order parameters. $NbSe_2$-Bi2212/Bi2201 heterojunctions constitute a unique platform for exploring emergent superconducting phenomena such as Ising superconductivity [23,24], anomalous metal state [25], and finite-momentum pairing [26,27].

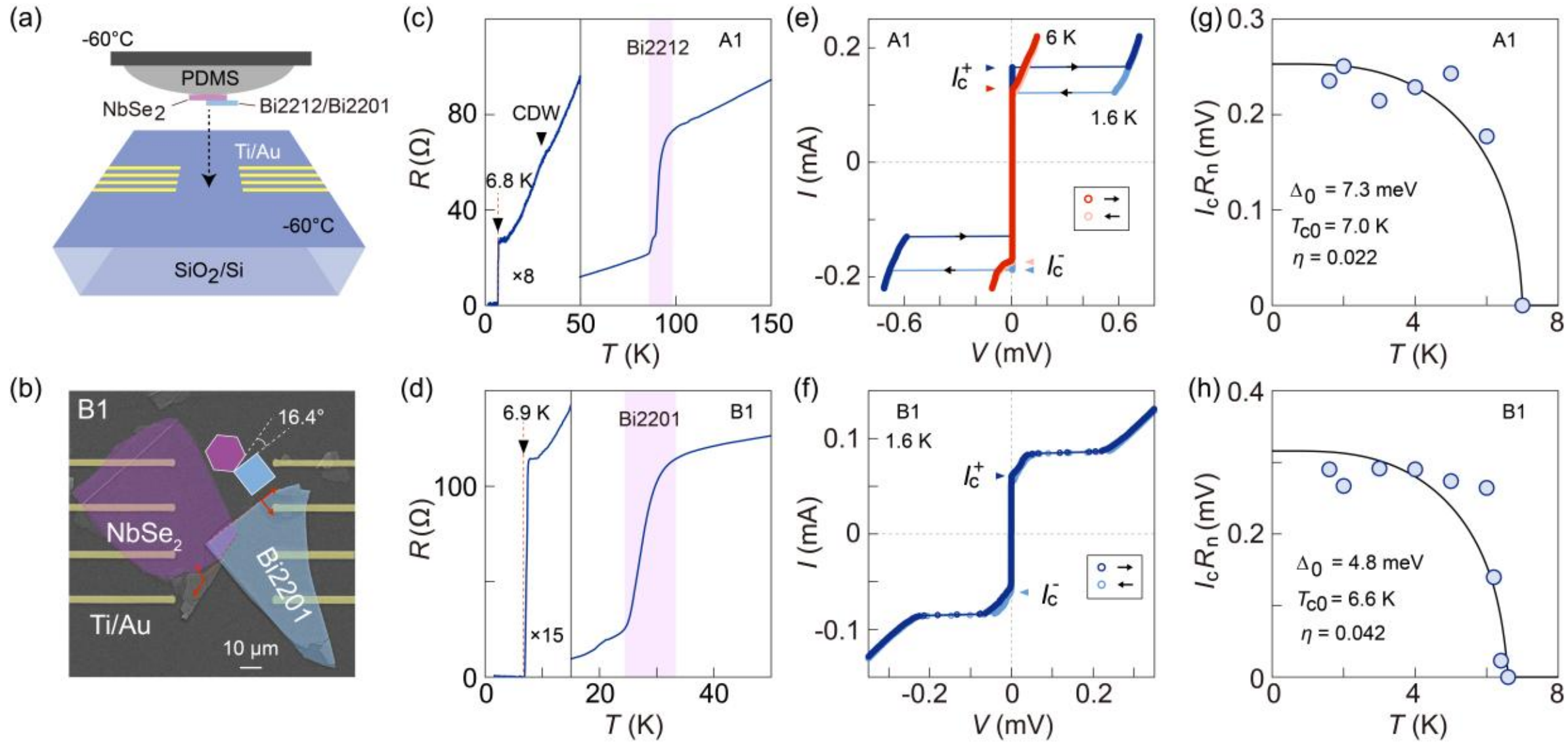

***Figure 1: Transport characteristics of $NbSe_2$-Bi2212 and $NbSe_2$-Bi2201 junctions.*** *(a) Schematic drawing of the van der Waals stacking process for making the hybrid junction. (b) Scanning electron microscopy image of the NbSe2-Bi2201 junction B1 (false colored). Crystalline orientations (indicated by arrows) are estimated based on the straight edges of the flakes and their relative angles. (c)(d) Temperature dependent junction resistances of sample A1 ($NbSe_2$-Bi2212) and B1 ($NbSe_2$-Bi2201). Resistance values in the low temperature regions are amplified for clarity. Arrows mark the superconducting transition temperature of the junction. (e)(f) $I$-$V$ characteristics across the junction of samples S1 and S5. Triangles mark the critical currents. Arrows indicate the sweeping directions. (g)(h) Temperature dependence of $I_cR_n$ of samples A1 and B1. Solid curves are theoretical fits by using the modified Ambegaokar-Baratoff formula: $I_cR_n = \eta \frac{\pi\Delta}{2e} tanh \frac{\Delta}{2k_BT}$, with $\Delta = \Delta_0\, tanh\left(1.74\sqrt{T_{c0}/T - 1}\right)$. Numbers in the panels show the fit parameters.*

We employ the all-cold stacking protocol to realize the heterojunctions of $NbSe_2$ and Bi2212/Bi2201 [Fig. 1(a)]. At a temperature of -60 °C, we use the mechanically exfoliated $NbSe_2$ flake on PDMS to pick up the Bi2212/Bi2201 flake after *in-situ* cleavage to expose a fresh surface. The whole stack is then transferred at -60 °C onto the $SiO_2$/Si substrate with prepatterned electrodes. Figure 1(b) shows the false-colored scanning electron microscope image of a typical device. The thicknesses of $NbSe_2$ are estimated to be about 50 to 200 nm, based on the optical contrast and atomic force microscopy. We estimate the twist angles by taking the straight edges of the single crystalline flakes. In Fig. 1(c)(d), we show typical temperature dependent resistance across the junctions. The sudden drop of resistance as the temperature decreases to around 80 or 20 K corresponds to the superconducting transition of the constituent Bi2212 or Bi2201 flake. We also observe a bump in the junction resistance around 30 K (marked by arrows), attesting to the onset of charge density wave in $NbSe_2$. Below 7 K, both $NbSe_2$-Bi2212 (A1) and $NbSe_2$-Bi2201 (B1) junctions show zero resistance. We have in total fabricated four $NbSe_2$-Bi2212 junctions and four $NbSe_2$-Bi2201 junctions. The critical temperature of these hybrid junctions is always governed by the superconducting transition temperature of $NbSe_2$.

Figure 1(e)(f) show the typical $I$-$V$ characteristics measured in a standard four-terminal configuration across the junctions. We observe a vertical branch at zero bias, indicating the Josephson tunneling. In sample A1, a $NbSe_2$-Bi2212 junction, sweeping the current above the critical value $I_c$ typically results in the sudden jump to finite voltage across the junction. The sudden jump with sweeping hysteresis is consistent with the behavior of a superconductor-insulator-superconductor junction. In contrast, sample B1 ($NbSe_2$-Bi2201) hosts a more gradual increase in voltage, reflecting high damping in the junction. In addition, we can observe a second voltage jump at a higher current, corresponding to the critical current in the $NbSe_2$ flake (Fig. S3 of the supplementary information). Interestingly, sample A1 shows asymmetry in the $I$-$V$ characteristics, indicating prominent superconducting diode effect [14]. The diode efficiency η, characterized as $\eta = \left| I_c^+ - |I_c^-| \right| / (I_c^+ + |I_c^-|)$, is 15% at 6 K. This value surpasses the typical value (around 10%) in $NbSe_2$-$NbSe_2$ homojunctions [28,29] and is comparable to that in the $NbSe_2$-$NbSe_2$ junction with intentionally introduced interface disorder [30]. It indicates that the hybrid junction is more susceptible to extrinsic effects such as trapped vortices induced by remanent magnetic fields. Figure 1(g)(h) show the corresponding temperature dependences of the Josephson coupling strength—the product of $I_c$ and the normal state resistance $R_n$—for A1 and B1. The obtained $I_c R_n$ is about 0.24 mV for A1 and 0.29 mV for B1. These values are orders or magnitude larger than those reported in earlier Pb-Bi2212 (1.2-7.7 μV) [20] and Nb-Bi2212 (1.8-4.5 μV) [21] junctions. They are comparable to the values in Pb-YBCO junctions [18]. The enhanced Josephson coupling suggests high interface quality of the van der Waals hybrid junctions.

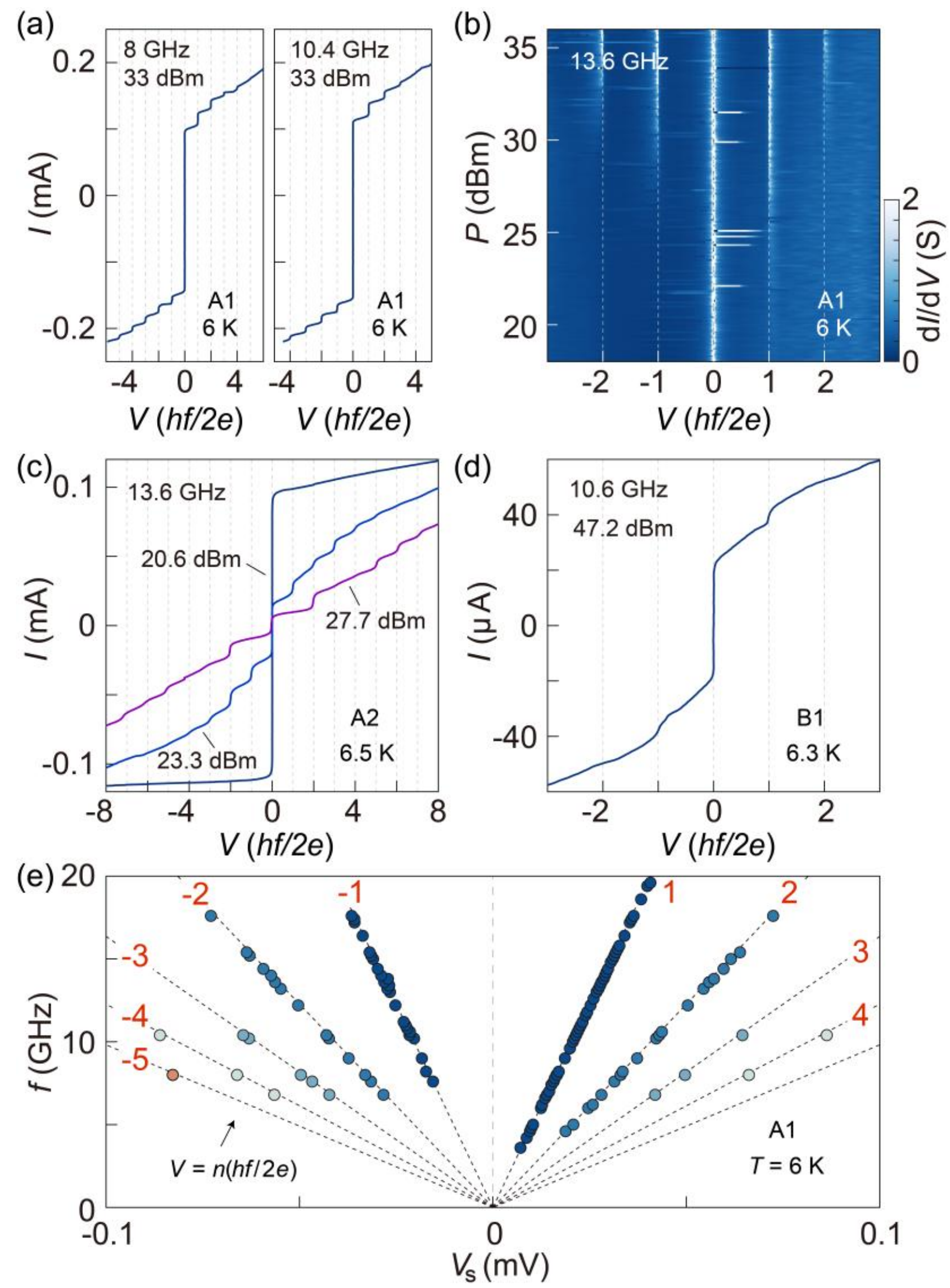

***Figure 2: Shapiro steps of both $NbSe_2$-Bi2212 and $NbSe_2$-Bi2201 junctions.*** *(a) $I$-$V$ characteristics of sample A1 under microwave irradiation. Numbers in the panel indicate the microwave frequency and power. (b) Color-coded plot of the junction conductance as a function of microwave power and junction voltage. (c) $I$-$V$ characteristics of sample A2 under microwave irradiation with different microwave powers. (d) $I$-$V$ characteristics of sample B1 under microwave irradiation. (e) Voltage positions of the Shapiro steps in sample A1 as a function of microwave frequency. Dashed lines represent the theoretical formula: $V_s = nhf/2e$, with $n$ being an integer.*

Figure 2 demonstrates the ac Josephson effect of the hybrid junctions. Under a sufficiently strong microwave irradiation, samples A1, B1, and A2 (another $NbSe_2$-

Bi2212 junction) show staircase structures in the $I$-$V$ characteristics [Fig. 2(a)(c)(d)], attesting to the emergence of Shapiro steps. Notably, the steps possess voltage positions only at integer times $hf/2e$, where $f$ is the microwave frequency, as indicated by the dashed vertical lines in Fig. 2(a)-(d). In Fig. 2(e), we further summarize the voltage positions in sample A1 a function of $f$. Clearly, the Shapiro steps follow the standard sequence of $nhf/2e$, where $n = \pm 1, \pm 2, \pm 3, \pm 4$. Both samples A1 and A2 show pronounced Josephson diode effect. Phenomenologically, such a diode effect requires the CPR of the junctions to host second harmonic term and time-reversal symmetry breaking term [31,32]. The absence of fractional Shapiro steps here indicates that the second harmonic term is still subdominant than the fundamental harmonic in the CPR, because otherwise half-integer Shapiro steps would emerge [10,14].

Figure 3 further confirms the Josephson nature of our heterojunctions by revealing the Fraunhofer diffraction pattern. Here, this sample (A3) hosts a relatively smaller critical current [Fig. 3(a)] than that of A1 and A2, possibly due to stronger oxygen out-diffusion of Bi2212 in this earlier batch of samples (We defer further discussion to the end of this letter). When the magnetic field is applied out-of-plane [Fig. 3(b)], we observe a quick decay of the critical current and the persistence of a suppressed critical current (about 5 μA) to a magnetic field as large as 1.5 T. On the other hand, applying the magnetic field in the plane—perpendicular to the direction of the tunneling current—results in the periodic modulation of Josephson current, as shown in Fig. 3(c). The first node of the Fraunhofer diffraction pattern occurs at $B_{\parallel}^{1} = \pm 0.1$ T. It corresponds to a junction area of about 0.02 μm$^2$, by employing the relation: $|B_{\parallel}^{1}|S = h/2e$. Furthermore, we evaluate the junction width to be 14 μm based on the optical image. Combing these quantities gives rise to a tunneling distance of about 1.4 nm, which is comparable to the thickness (1.5 nm) for half a unit cell of Bi2212. In reality, the tunneling current may not spread uniformly across the complete overlapping area between $NbSe_2$ and Bi2212 due to interface inhomogeneity [20]. This deviation also manifests itself in the Fraunhofer diffraction pattern. For a junction with a uniform distribution of the tunneling current, the second node is expected to appear at around $\pm 0.2$ T, as illustrated by the solid curves. In contrast, the experimental results show suppression of the second lobe around $\pm 0.3$ T. The dashed curves in Fig. 3(c) better capture this delayed suppression. The corresponding current distribution is shown in the inset. It indicates that the tunneling current is suppressed in the center of the junction, akin to the behavior previously observed in one of the Pb-Bi2212 junctions [20].

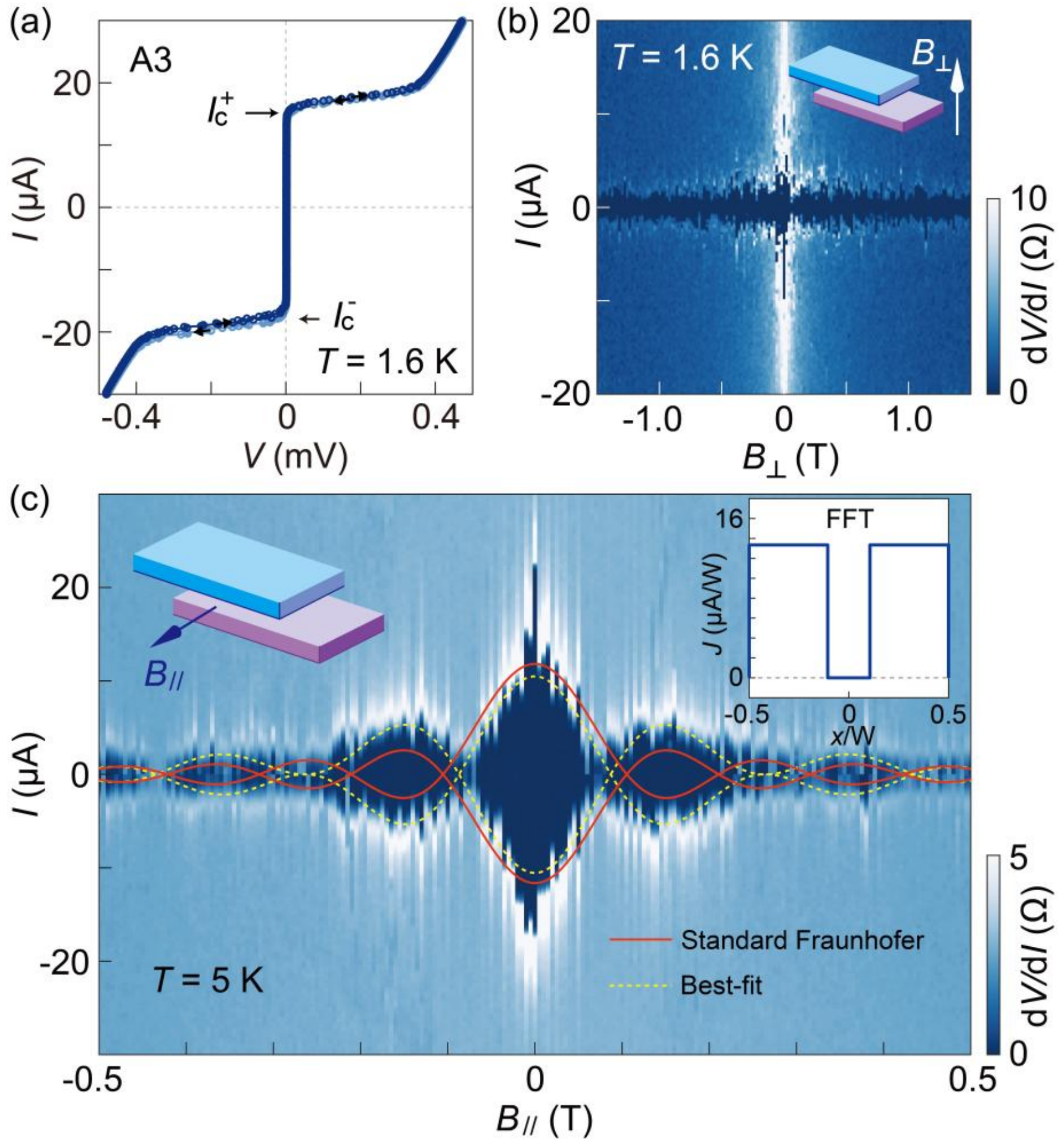

***Figure 3: Fraunhofer diffraction pattern of the $NbSe_2$-Bi2212 Josephson junction.*** *(a) $I$-$V$ characteristics of sample A3. (b) Colored plot of the junction resistance ($dV/dI$) as a function of the bias current and the perpendicular magnetic field. Inset illustrates the orientation of the magnetic field relative to the junction. (c) Colored plot of the junction resistance ($dV/dI$) as a function of the bias current and the in-plane magnetic field (schematically shown as the upper-left inset). Solid and dashed curves indicate the theoretically calculated Fraunhofer diffraction patterns, assuming a uniform current distribution (solid) and non-uniform one (dashed, shown as the upper-right inset), respectively.*

In general, we experimentally demonstrate that the Josephson coupling can be achieved between $NbSe_2$ and Bi2212/Bi2201. To understand these results, we theoretically calculate the Josephson coupling. We consider the quasi-two-dimensional bands of the two materials as plotted in Fig. 4(a) [33,34]. We adopt the formula developed for twisted cuprate Josephson junctions [2,6]:

$$J_c(\theta) = \left|4eT \sum_{\omega} \int_{BZ1} \int_{BZ2} F_{1,2}\left(\overrightarrow{k_{1,r}}, \overrightarrow{k_2}\right) G_{\omega}^{1}\left(\overrightarrow{k_1}\right) G_{\omega}^{2}\left(\overrightarrow{k_2}\right)\right|. \quad (1)$$

Here, we sum up the Matsubara frequency and integrate over the first Brillouin zones (BZ1 and BZ2) of $NbSe_2$ and Bi2212 with their in-plane momenta $\overrightarrow{k_1}$ and $\overrightarrow{k_2}$. We use $F_{1,2} = F_0\, exp\left[-\left(\overrightarrow{k_{1,r}} - \overrightarrow{k_2}\right)^2/(2\pi^2\sigma^2)\right]$ to describe tunneling under varying degrees of in-plane momentum conservation. A fully incoherent tunneling corresponds to $\sigma \to \infty$. $\overrightarrow{k_{1,r}}$ represents the in-plane momentum that is rotated from the original $\overrightarrow{k_1}$ by a twist angle of $\theta$. Here, $\theta = 0$ corresponds to the case where the $\Gamma$-K direction of $NbSe_2$ is parallel to $\Gamma$-X of Bi2212/Bi2201. The anomalous Green functions follow $G_{\omega}^{1,2} = \Delta_{1,2}/\left(\omega^2 + \xi_{1,2}^2 + \Delta_{1,2}^2\right)$, where $\Delta_{1,2}$ represents the momentum dependent superconducting gaps of the two materials and $\xi_{1,2}$ corresponds to the quasiparticle dispersions.

Figure 4(b) shows the calculated Josephson current as a function of $\theta$ for $\sigma = 0.1$. In this case, the tunneling coherence is largely preserved such that crystalline symmetries play an important role. All the calculated angular dependences inherit the six-fold rotational symmetry from $NbSe_2$, as indicated by the $60°$-period of the curves for a mixed order parameter in Bi2212/Bi2201. In the case when Bi2212/Bi2201 is assigned to a pure *d*-wave order parameter, the four-fold symmetry of pure *d*-wave pairing further reduces the period to $30°$. Between 0 and $30°$, the Josephson tunneling is generally non-zero. It is suppressed only at a specific twist angle of $15°$, if the two order parameters possess a pure *d*-wave form [$\Delta_1 = \Delta_d\left(cos\, k_{x,2} - cos\, k_{y,2}\right)$] and a pure *s*-wave one ($\Delta_2 = \Delta_s$).

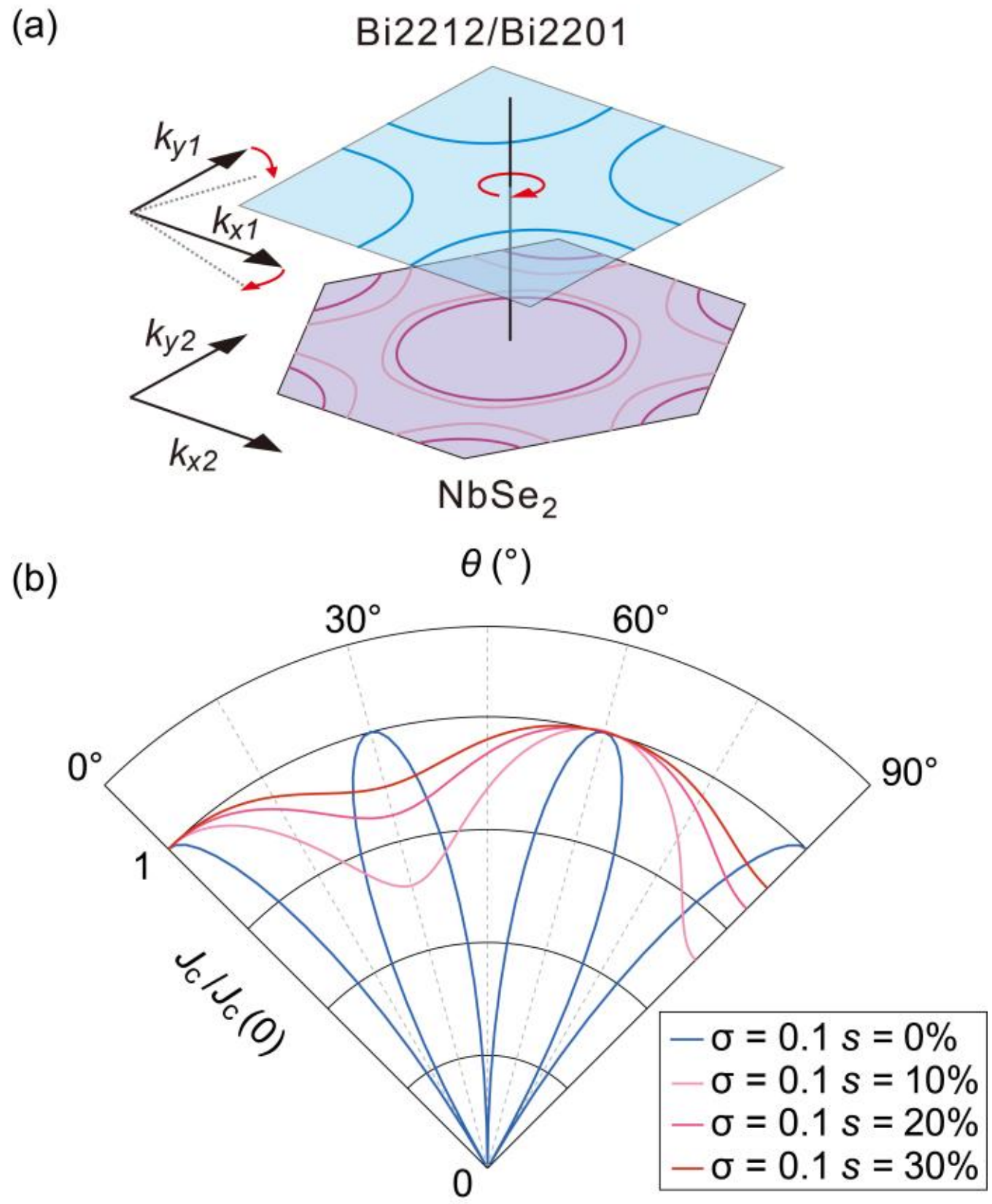

**Figure 4: Theoretical analysis of the hybrid junction.** (a) Illustration of the Fermi-level crossing bands in the first Brillouin zones of Bi2212/Bi2201 and $NbSe_2$, respectively. Zero twist angle corresponds to the situation that $k_{x1}$ is parallel to $k_{x2}$. (b) Theoretically calculated Josephson current as a function of the twist angle (normalized by the value at zero twist angle).

The theoretical calculations indicate that the hybrid junctions constitute an alternative platform to the twisted cuprate junctions for addressing the pairing symmetry. In particular, the comparison of the Josephson coupling strengths at $0°$ and $15°$ would be a phase-sensitive detection scheme. A systematic investigation in this direction calls for future endeavors. Still, we summarize in Tab. 1 the data points from the samples of this work. We list the Josephson critical current densities obtained in our hybrid junctions, together with the twist angles evaluated for samples A1, A3, B1, B3, and B4. Sample B1 is of particular interest because its twist angle is close to $15°$, at which the coherent Josephson tunneling is expected to be fully suppressed between a pure *d*-wave superconductor and a pure *s*-wave superconductor. In contrast, the critical current density of B1 is comparable to those obtained from B3 ($25°$) and B4 (29°). This angular insensitivity seems to favor an emergent $s$-wave component in Bi2201 [8]. Still, a stringent test on the pairing symmetry would require meticulous control over the sample quality. The $NbSe_2$-Bi2212 samples show a large variation of $J_c$. $J_c$ in A4 is nearly two orders of magnitude as large as that of A3, although A3 has an estimated twist angle of 55 degrees, at which $J_c$ is theoretically expected to reach maximum. Apart from the variation in the transition temperature of the junction, we attribute the difference in $J_c$ to different extent of oxygen out-diffusion on the Bi2212 side during the sample fabrication process. The stronger oxygen out-diffusion (especially in the earlier batch of samples with extended sample fabrication time) may cause higher inhomogeneity in the Bi2212 layer that is adjacent to $NbSe_2$. The temperature dependent resistances reflect this variation. For instance, the normal state resistance of device A3 slightly deviates from a linear temperature dependent behavior (Fig. S2 of the supplementary information), whereas the normal state resistance of device A4 nicely agrees with the standard linear temperature dependence. In contrast, Bi2201 samples show weaker variation, possibly suggesting that oxygen out-diffusion is less severe in this La-doped compound [8].

Finally, we envision that future experiments with thinner $NbSe_2$ can help address emergent phenomena arising from the interplay between reduced dimension and strong spin-orbit coupling [23-27]. Especially, the Josephson tunneling can be a sensitive probe to the first-order phase transition of the superconducting gap of 20-30 nm thick $NbSe_2$, driven by the in-plane magnetic field [27]. Furthermore, our technique is suitable for a broad class of hybrid Josephson junctions, including Kagome superconductor-cuprate superconductor junctions, iron-based superconductor-cuprate superconductor junctions, etc. Constructing those hybrid junctions offers a key advantage. The charge reservoir layer of Bi2212 or Bi2201 itself serves as an ultra-thin, well-defined insulating barrier. It eliminates the contamination and degradation that may be associated with introducing a separate insulating material into the junction.

**Table I. Summary of the hybrid junctions studied in this work.** List of samples and their corresponding transition temperatures $T_{c0}$, critical current densities $J_c$, and estimated twist angles. We calculate $J_c$ by dividing the measured critical current by the physical area of the junction, as determined from optical images. We denote $NbSe_2$-Bi2212 ($NbSe_2$-Bi2201) samples as A1, A2, … (B1, B2, …). Samples without identifiable straight edges (A2, A4, B2) prevent us from determining the twist angles.

| Device | $T_{c0}$ (K) | $J_c$ (A/$cm^2$) @ 1.6 K | Twist angle (°) |
|---|---|---|---|
| A1 | 6.8 | 123 | 54 |
| A2 | 7.0 | 582 | - |
| A3 | 6.8 | 10 | 55 |
| A4 | 6.8 | 893 | - |
| B1 | 6.8 | 110 | 16 |
| B2 | 6.8 | 34 | - |
| B3 | 6.9 | 197 | 69 |
| B4 | 6.9 | 200 | 65 |

## Acknowledgements

We thank Genda Gu for providing the Bi2212 crystals. This work is financially supported by the National Natural Science Foundation of China [Grants No. 52388201 (QKX), T2425009 (DZ), 12141402 (YZ), 92477101 (YZ), 12488201 (XZ)]; Ministry of Science and Technology of China [Grant No. 2024YFA1408101 (YZ), 2022YFA1403100 (DZ, QKX), 2021YFA1401800 (XZ)]; Innovation Program for Quantum Science and Technology [Grants No. 2021ZD0302600 (YZ), 2021ZD0302400 (DZ)].

## Author contributions

Y. Z. fabricated the devices. Y. Z. and Z. L. carried out the transport measurements. H. Y., L. Z., and X. J. Z. grew the single crystals of Bi2201. Y. Z., Z. L., Q.-K. X., and D.Z. analyzed the data and wrote the paper. All authors discussed the results and commented on the manuscript.